\documentclass[12pt]{article}

\usepackage{graphicx}
\usepackage{amsmath,amssymb}
\usepackage{bm}
\usepackage{graphicx, color}
\usepackage{wrapfig}
\usepackage{cancel}
\begin{document}
\title{
\begin{flushright}
\ \\*[-80pt] 
\begin{minipage}{0.2\linewidth}
\normalsize
HUPD1604 \\
KIAS-P16033 \\*[50pt]
\end{minipage}
\end{flushright}
{\Large \bf 
$A_4 \times U(1)_{PQ}$ Model for the Lepton Flavor Structure \\
and the Strong $CP$ Problem
\\*[20pt]}}

\author{ 
\centerline{
Takaaki~Nomura$^{1,}$\footnote{E-mail address: nomura@kias.re.kr},
~Yusuke~Shimizu$^{1,2}$\footnote{E-mail address: yu-shimizu@hiroshima-u.ac.jp},~and
~Toshifumi~Yamada$^{1,}$\footnote{E-mail address: toshifumi@kias.re.kr} }
\\*[20pt]
\centerline{
\begin{minipage}{\linewidth}
\begin{center}
$^1${\it \normalsize School~of~Physics,~KIAS,~Seoul~130-722,~Republic~of~Korea} \\*[10pt]
$^2${\it \normalsize Graduate~School~of~Science,~Hiroshima~University, \\
Higashi-Hiroshima,~739-8526,~Japan}
\end{center}
\end{minipage}}
\\*[50pt]}

\date{
\centerline{\small \bf Abstract}
\begin{minipage}{0.9\linewidth}
\medskip 
\medskip 
\small
We present a model with $A_4 \times U(1)_{PQ}$ lepton flavor symmetry which explains the origin of the lepton flavor structure and also solves the strong $CP$ problem.
Standard model gauge singlet fields, so-called ``flavons", charged under the $A_4 \times U(1)_{PQ}$ symmetry are introduced and are coupled with the lepton and the Higgs sectors.
The flavon vacuum expectation values (VEVs) trigger spontaneous breaking of the $A_4 \times U(1)_{PQ}$ symmetry.
The breaking pattern of the $A_4$ accounts for the tri-bimaximal neutrino mixing and the deviation from it due to the non-zero $\theta_{13}$ angle,
 and the breaking of the $U(1)_{PQ}$ gives rise to a pseudo-Nambu-Goldstone boson, axion, whose VEV cancels the QCD $\theta$ term.
We investigate the breaking of the $A_4 \times U(1)_{PQ}$ symmetry through an analysis on the scalar potential
 and further discuss the properties of the axion in the model, including its decay constant, mass and coupling with photons.
It is shown that the axion decay constant is related with the right-handed neutrino mass through the flavon VEVs.
Experimental constraints on the axion and their implications are also studied.
\end{minipage}
}

\begin{titlepage}
\maketitle
\thispagestyle{empty}
\end{titlepage}

\section{Introduction}
The standard model (SM) is the most attractive model which explains almost all the results of collider and low-energy experiments,
 and the last piece of the SM, the Higgs particle, has been discovered very recently.
However there remains a mystery about the origin of the structures of quark and lepton flavors.
As to the lepton sector, neutrino oscillation experiments have provided interesting information about the flavor structure in the form of
the two neutrino mass squared differences and the two large mixing angles. 
The reactor neutrino experiments~\cite{An:2012eh}-\cite{Abe:2014lus} have also reported a non-zero $\theta _{13}$, 
which was the last mixing angle to be measured in the lepton sector.
Deriving the neutrino mass and mixing angles theoretically is a challenge in the quest of physics beyond the SM. 
The non-Abelian discrete symmetry is one of the candidates for the origin of the flavor structures. 
Many authors have proposed models with non-Abelian discrete flavor symmetries in the lepton sector as well as in the quark sector 
(See for review~\cite{Ishimori:2010au}-\cite{King:2014nza}.).
Those models contain new gauge singlet scalar fields, so-called ``flavons", in addition to the $SU(2)$ doublet SM Higgs field. 
In order to obtain vacuum expectation values (VEVs) and VEV alignments of flavon scalar fields, 
one usually introduces so-called ``driving fields" in the framework of the supersymmetry (SUSY) with $U(1)_R$ symmetry.

The strong $CP$ problem is another mystery of the SM.
In QCD theory, a non-zero QCD vacuum angle $\theta$ is allowed which leads to the violation of $CP$ symmetry.
Searches for neutron electric dipole moment report an experimental bound of $\vert \theta \vert \lesssim 10^{-11}$~\cite{Baker:2006ts}.
Such an unnaturally small $\theta$ requires a theoretical explanation.
The most elegant solution to the strong $CP$ problem is to introduce the $U(1)_{PQ}$ symmetry~\cite{Peccei:1977hh}.
The $U(1)_{PQ}$ symmetry is anomalous in QCD, and when spontaneously broken, it yields a pseudo-Nambu-Goldstone boson called axion.
The axion field develops a VEV that cancels the $\theta$ term.
The axion decay constant, which is the scale of $U(1)_{PQ}$ symmetry breaking, 
should be much higher than the electroweak scale to evade the constraint from supernova cooling~\cite{1987a}.
Hence the breaking of $U(1)_{PQ}$ symmetry is associated with the VEV of some new scalar field with a $U(1)_{PQ}$ charge.
The pervasive models of the axion are KSVZ model \cite{ksvz} and DFSZ model \cite{dfsz}.
In the former, the new scalar couples with new vector-like quarks, while in the latter, two Higgs doublets are introduced and are coupled with the new scalar.

In this paper, we present a model with $A_4 \times U(1)_{PQ}$ symmetry that explains the origin of the lepton flavor structure and also solves the strong $CP$ problem.
We show that one can successfully assign $A_4$ and $U(1)_{PQ}$ charges to the leptons, quarks, Higgs fields and flavons.
Note that the quarks and Higgs fields are $A_4$ trivial singlets, which allows the same quark Yukawa couplings as in the minimal SUSY SM (MSSM).
$A_4$ flavor symmetry regulates the flavor structure of the lepton sector.
Since flavon fields have $U(1)_{PQ}$ charges, non-zero VEVs of flavons realize the spontaneous breaking of $U(1)_{PQ}$ symmetry, yielding an axion field.
The Higgs fields have $U(1)_{PQ}$ charges and the axion in our model is of DFSZ-type.
What is unique in our model is that the axion decay constant is tied with flavon VEVs, which are also related with the right-handed Majorana neutrino mass.
Another appeal is that, in conventional $A_4$ flavor symmetry models, additional $Z_N$ symmetry is often imposed to forbid certain terms, 
while in our model, the $U(1)_{PQ}$ symmetry plays the role of the $Z_N$ symmetry and no ad-hoc $Z_N$ symmetry is necessary.
In Ref.~\cite{Ahn:2014gva}, another model with $A_4\times U(1)_{PQ}$ symmetry has been presented. 
There, flavon fields with $U(1)_{PQ}$ charges couple directly with the quark sector and the axion has a property different from the DFSZ and KSVZ axions.
In our model, on the other hand, they couple to quarks only through the two Higgs fields and the axion is the DFSZ axion.
Hence the two models are distinctively different.
\footnote{
In Ref.~\cite{Antusch:2013rla}, a different type of model has been presented which explains the quark flavor structure with discrete flavor symmetries 
and also solves the strong $CP$ problem. 
Unlike our model, that model's solution to the strong $CP$ problem is based on the idea 
that $CP$ is a fundamental symmetry of Nature and the $CP$ phase of the SM is provided by spontaneous $CP$ symmetry breaking.
}

This paper is organized as follows. In section~\ref{sc:model}, we present our model of $A_4\times U(1)_{PQ}$ flavor symmetry with SUSY and $U(1)_R$ symmetry,
and discuss the potential of the flavon and driving fields. We also comment on the origin of the Higgs $\mu$-term in the model.
In section~\ref{sc:axion}, we study the properties of the axion in the model and its connection to the flavon properties.
Also, experimental constraints on the axion are discussed.
The section~\ref{sc:summary} is devoted to summary. In appendix~\ref{sec:multiplication-rule}, 
we review the multiplication rule of $A_4$ group.



\section{$A_4\times U(1)_{PQ}$ model}
\label{sc:model}
We construct an $A_4\times U(1)_{PQ}$ model for describing the quark and lepton sectors. 
In this model, we require that the left-handed lepton doublets $l=(l_e, l_\mu, l_\tau)$ transform as a triplet under the $A_4$,
while the right-handed charged leptons are assigned to the singlets as $\bf 1$, $\bf 1''$, and $\bf 1'$ for $e_R^c$, $\mu _R^c$, and $\tau _R^c$. 
The right-handed Majorana neutrinos are introduced as an $A_4$ triplet $\nu ^c_R=(\nu _{eR}^c, \nu _{\mu R}^c,\nu ^c_{\tau R})$. 
On the other hand, the left-handed quark doublets $q_1,q_2,q_3$ and the right-handed up- and down-type quarks $u_R^c,c_R^c,t_R^c,d_R^c,s_R^c,b_R^c$ transform as trivial singlets under the $A_4$. 
For the Higgs sector, we introduce the Higgs doublets $h_u$ and $h_d$ which are assigned to $A_4$ trivial singlets.
The setup accommodates the same Yukawa couplings for the quarks as in MSSM.
\begin{table}[hb]
\begin{center}
\begin{tabular}{|c||c|ccc|c|ccc|ccc|ccc|}
\hline
& $l$ & $e_R^c$ & $\mu _R^c$ & $\tau _R^c$ & $\nu _R^c$ & $q_1$ & $q_2$ & $q_3$ & $u_R^c$ & $c_R^c$ & $t_R^c$ & $d_R^c$ & $s_R^c$ & $b_R^c$ \\ 
\hline 
$SU(2)$ & $2$ & \multicolumn{3}{c|}{$1$} & $1$ & \multicolumn{3}{c|}{$2$} & \multicolumn{3}{c|}{$1$} & \multicolumn{3}{c|}{$1$} \\
$A_4$ & $\bf 3$ & $\bf 1$ & $\bf 1''$ & $\bf 1'$ & $\bf 3$ & \multicolumn{3}{c|}{$\bf 1$} & \multicolumn{3}{c|}{$\bf 1$} & \multicolumn{3}{c|}{$\bf 1$} \\
$U(1)_{PQ}$ & $\frac{1}{4}\left (1-4q_u\right )$ & \multicolumn{3}{c|}{$\frac{1}{4}\left (8q_u-5\right )$} & $-\frac{1}{4}$ 
& \multicolumn{3}{c|}{$r$} & \multicolumn{3}{c|}{$-q_u-r$} & \multicolumn{3}{c|}{$q_u-r-1$} \\
$U(1)_R$ & $1$ & \multicolumn{3}{c|}{$1$} & $1$ & \multicolumn{3}{c|}{$1$} & \multicolumn{3}{c|}{$1$} & \multicolumn{3}{c|}{$1$} \\
\hline 
\end{tabular}

\vspace{1mm}
\begin{tabular}{|c||cc||c|ccc||c|cc|}
\hline
& $h_u$ & $h_d$ & $\phi _T $ & $\phi _S$ & $\xi $ & $\xi '$ & $\phi _0^T$ & $\phi _0^S$ & $\xi _0$ \\ 
\hline 
$SU(2)$ & \multicolumn{2}{c||}{$2$} & $1$ & \multicolumn{3}{c||}{$1$} & $1$ & \multicolumn{2}{c|}{$1$} \\
$A_4$ & \multicolumn{2}{c||}{$\bf 1$} & $\bf 3$ & $\bf 3$ & $\bf 1$ & $\bf 1'$ & $\bf 3$ & $\bf 3$ & $\bf 1$ \\
$U(1)_{PQ}$ & $q_u$ & $(1-q_u)$ & $0$ & \multicolumn{3}{c||}{$\frac{1}{2}$} & $0$ & \multicolumn{2}{c|}{$-1$} \\
$U(1)_R$ & \multicolumn{2}{c||}{$0$} & $0$ & \multicolumn{3}{c||}{$0$} & $2$ & \multicolumn{2}{c|}{$2$} \\
\hline 
\end{tabular}
\end{center}
\caption{The assignments of leptons, quarks, Higgs, flavons, and driving fields. 
The $U(1)_{PQ}$ charges are normalized in such a way that the charges of $h_u$ and $h_d$ sum to 1.
$q_u$ and $r$ are arbitrary constants.}
\label{tab:assignment}
\end{table}
The gauge singlet flavons $\phi _T$, $\phi _S$, $\xi $, and $\xi '$ are added where
$\phi _T=(\phi _{T1},\phi _{T2},\phi _{T3})$ and $\phi _S=(\phi _{S1},\phi _{S2},\phi _{S3})$ are triplets, 
$\xi$ is trivial singlet, and $\xi '$ is singlet-prime under the $A_4$ symmetry, respectively. 
We introduce a $U(1)_{PQ}$ symmetry to forbid irrelevant couplings in the lepton sector. 
The $A_4\times U(1)_{PQ}$ charge assignments are shown in Table~\ref{tab:assignment},
where the $U(1)_{PQ}$ charges are normalized in such a way that the sum of the charges of the Higgs doublets is $1$.
In addition, we introduce the so-called ``driving fields" $\phi _0^T=(\phi _{01}^T,\phi _{02}^T,\phi _{03}^T)$ and 
$\phi _0^S=(\phi _{01}^S,\phi _{02}^S,\phi _{03}^S)$ which are $A_4$ triplets, and $\xi _0$ which is $A_4$ trivial 
singlet, in order to obtain VEVs and VEV alignments for the flavons. 
Then the VEV alignments can be generated through $F$-terms which couples flavons to driving fields and 
carries the $R$ charge $+2$ under $U(1)_R$ symmetry. For leptons and quarks, we also assign $R$ charge $+1$. 
The charge assignments of driving fields are also shown in Table~\ref{tab:assignment}.
In the setup, the superpotential $w$ 
respecting $A_4 \times U(1)_{PQ}$ symmetry 
at the leading order is written as~\footnote{This model is same charge assignments of Ref.~\cite{Muramatsu:2016bda} 
except for $Z_3$ charges because we introduce $U(1)_{PQ}$ symmetry instead of $Z_3$ symmetry.} 
\begin{align}
\label{eq:superpotential}
w&\equiv w_d +w_Y, \nonumber \\
w_d&\equiv w_d^T+w_d^S, \nonumber \\
w_d^T&=-M\phi _0^T\phi _T+g\phi _0^T\phi _T\phi _T, \nonumber \\
w_d^S&=g_1\phi _0^S\phi _S\phi _S -g_2\phi _0^S\phi _S\xi +g_2'\phi _0^S\phi _S\xi '+g_3\xi _0\phi _S\phi _S-g_4\xi _0\xi \xi +\lambda \xi _0h_uh_d~, \nonumber \\ 
w_Y&\equiv w_\ell +w_D+w_N+w_{Y_u}+w_{Y_d}, \nonumber \\
w_\ell &=y_e\phi _Tle_R^ch_d/\Lambda +y_\mu \phi _Tl\mu _R^ch_d/\Lambda +y_\tau \phi _Tl\tau _R^ch_d/\Lambda , \nonumber \\
w_D&=y_Dl\nu _R^ch_u, \nonumber \\
w_N&=y_{\phi _S}\phi _S\nu _R^c\nu _R^c+y_\xi \xi \nu _R^c\nu _R^c+y_{\xi '}\xi '\nu _R^c\nu _R^c, \nonumber \\
w_{Y_u}&=y_{i\alpha }q_i\alpha _R^ch_u~(i=1,2,3,~\alpha =u,c,t), \nonumber \\
w_{Y_d}&=y_{i\beta }q_i\beta _R^ch_d~(i=1,2,3,~\beta =d,s,b),
\end{align}
where $M$ is generally complex mass parameter, $g$'s are trilinear couplings which are also complex parameters
\footnote{In order to obtain the positive number of $v_T$, $v_S$, $u$, and $u'$ for Eqs.~(\ref{eq:alignment-vT}) and (\ref{eq:alignment-vS}),  
we take negative sign for several terms in Eq.~(\ref{eq:superpotential}).}, 
$\lambda $ is trilinear coupling for $SU(2)$ doublet Higgs and driving field, 
$y'$s are complex Yukawa couplings, 
and $\Lambda$ is the $A_4$ cutoff scale. 
Note that the above charge assignment is consistent with the quark sector of the MSSM, since we assign the quark fields to $A_4$ trivial singlets.
From this superpotential, we discuss the potential analysis including flavons and driving fields in the next subsection.

\subsection{Potential analyses including flavons and driving fields}
Let us discuss the potential for scalar fields including flavons and driving fields. 
The superpotential $w_d^T$ and $w_d^S$ in Eq.~(\ref{eq:superpotential}) are obtained as 
\begin{align}
w_d^T&=-M\phi _0^T\phi _T+g\phi _0^T\phi _T\phi _T \nonumber \\
&=-M\left (\phi _{01}^T\phi _{T1}+\phi _{02}^T\phi _{T3}+\phi _{03}^T\phi _{T2}\right ) \nonumber \\
&+\frac{2g}{3}\Big [\phi _{01}^T\left (\phi _{T1}^2-\phi _{T2}\phi _{T3}\right )+\phi _{02}^T\left (\phi _{T2}^2-\phi _{T1}\phi _{T3}\right )
+\phi _{03}^T\left (\phi _{T3}^2-\phi _{T1}\phi _{T2}\right )\Big ]~, \nonumber \\
w_d^S&=g_1\phi _0^S\phi _S\phi _S -g_2\phi _0^S\phi _S\xi +g_2'\phi _0^S\phi _S\xi ' +g_3\xi _0\phi _S\phi _S-g_4\xi _0\xi \xi +\lambda \xi _0h_uh_d \nonumber \\
&=\frac{2g_1}{3}\Big [\phi _{01}^S\left (\phi _{S1}^2-\phi _{S2}\phi _{S3}\right )+\phi _{02}^S\left (\phi _{S2}^2-\phi _{S1}\phi _{S3}\right )
+\phi _{03}^S\left (\phi _{S3}^2-\phi _{S1}\phi _{S2}\right )\Big ] \nonumber \\
&-g_2\left (\phi _{01}^S\phi _{S1}+\phi _{02}^S\phi _{S3}+\phi _{03}^S\phi _{S2}\right )\xi 
+g_2'\left (\phi _{01}^S\phi _{S3}+\phi _{02}^S\phi _{S2}+\phi _{03}^S\phi _{S1}\right )\xi '\nonumber \\
&+g_3\xi _0\left (\phi _{S1}^2+2\phi _{S2}\phi _{S3}\right )-g_4\xi _0\xi ^2+\lambda \xi _0h_uh_d~.
\end{align}
In order to discuss the VEVs and VEV alignments of flavons, we consider 
the scalar potential except for driving fields, which is given as 
\begin{align}
V&\equiv V_T+V_S, \nonumber \\
V_T&=\sum _i\left |\frac{\partial w_d^T}{\partial \phi _{0i}^T}\right |^2+h.c. \nonumber \\
&=2\left |-M\phi _{T1}+\frac{2g}{3}\left (\phi _{T1}^2-\phi _{T2}\phi _{T3}\right )\right |^2
+2\left |-M\phi _{T3}+\frac{2g}{3}\left (\phi _{T2}^2-\phi _{T1}\phi _{T3}\right )\right |^2 \nonumber \\
&+2\left |-M\phi _{T2}+\frac{2g}{3}\left (\phi _{T3}^2-\phi _{T1}\phi _{T2}\right )\right |^2, \nonumber \\
V_S&=\sum \left |\frac{\partial w_d^S}{\partial X}\right |^2+h.c.\quad (X=\phi _{0i}^S,~\xi _0) \nonumber \\
&=2\left |\frac{2g_1}{3}\left (\phi _{S1}^2-\phi _{S2}\phi _{S3}\right )-g_2\phi _{S1}\xi +g_2'\phi _{S3}\xi '\right |^2
+2\left |\frac{2g_1}{3}\left (\phi _{S2}^2-\phi _{S1}\phi _{S3}\right )-g_2\phi _{S3}\xi +g_2'\phi _{S2}\xi '\right |^2 \nonumber \\
&+2\left |\frac{2g_1}{3}\left (\phi _{S3}^2-\phi _{S1}\phi _{S2}\right )-g_2\phi _{S2}\xi +g_2'\phi _{S1}\xi '\right |^2 
+2\left |g_3\left (\phi _{S1}^2+2\phi _{S2}\phi _{S3}\right )-g_4\xi ^2+\lambda h_uh_d\right |^2~.
\label{eq:scalar-potential}
\end{align}
Applying the condition of the potential minimum ($V_T=0$), we derive the VEV alignment of $\phi_T$ as 
\begin{equation}
\langle \phi _T\rangle =v_T(1,0,0),\qquad v_T=\frac{3M}{2g},
\label{eq:alignment-vT}
\end{equation}
where $v_T$ is generally complex number since $M$ and $g$ are complex.
By using VEV and VEV alignment of Eq.~(\ref{eq:alignment-vT}), 
we obtain diagonal mass matrix for the charged leptons.
Then the phase of $v_T$ can be removed and we
take $M/g$ as real parameter without loss of generality. 
In the following analysis, we adopt $M$ and $g$ as real parameters for simplicity. 
On the other hand, we derive the VEV alignment of $\phi _S$ and VEVs of $\xi $, $\xi '$, $h_u$ and $h_d$ 
from the condition of the potential minimum ($V_S=0$) in Eq.(\ref{eq:scalar-potential}) as
\begin{align}
\langle h_u\rangle =\langle h_d\rangle =0,\quad 
&\langle \xi \rangle =u,\quad \langle \xi '\rangle =u', \nonumber \\
\langle \phi _S\rangle =v_S(1,1,1),\quad &v_S^2=\frac{g_4}{3g_3}u^2,\quad u'=\frac{g_2}{g_2'}u.
\label{eq:alignment-vS}
\end{align}
As a result, we can take VEVs $u$ and $u'$ as free parameters.
Applying the VEVs and VEV alignment of Eq.~(\ref{eq:alignment-vS}), the neutrino mass matrix induces the lepton mixing 
as $1$-$3$ rotation from tri-bimaximal mixing (TBM)~\cite{Harrison:2002er,Harrison:2002kp}.
Furthermore, we obtain the non-zero $\theta _{13}$ which comes from $A_4$ singlet flavon VEV ratio $u'/u$ (See Refs.~\cite{Brahmachari:2008fn}-\cite{Karmakar:2014dva}.).

On the other hand the scalar potential including driving fields is given as
\begin{align}
\label{eq:scalar-potential-driving}
V_d&\equiv V_d^T+V_d^S, \nonumber \\
V_d^T&=\sum _i\left |\frac{\partial w_d^T}{\partial \phi _{Ti}}\right |^2+h.c. \nonumber \\
&=2\left |\phi _{01}^T\left (-M+\frac{4g}{3}\phi _{T1}\right )-\frac{2g}{3}\left (\phi _{02}^T\phi _{T3}+\phi _{03}^T\phi _{T2}\right )\right |^2 \nonumber \\
&+2\left |-\phi _{03}^T\left (M+\frac{2g}{3}\phi _{T1}\right )+\frac{2g}{3}\left (-\phi _{01}^T\phi _{T3}+2\phi _{02}^T\phi _{T2}\right )\right |^2 \nonumber \\
&+2\left |-\phi _{02}^T\left (M+\frac{2g}{3}\phi _{T1}\right )+\frac{2g}{3}\left (-\phi _{01}^T\phi _{T2}+2\phi _{03}^T\phi _{T3}\right )\right |^2, \nonumber \\
V_d^S&=\sum _i\left |\frac{\partial w_d^S}{\partial X_i}\right |^2+h.c.\quad \left (X_i=\phi _{Si},~\xi ,~\xi '\right ) \nonumber \\
&=2\left |\phi _{01}^S\left (\frac{4g_1}{3}\phi _{S1}-g_2\xi \right )-\frac{2g_1}{3}\phi _{02}^S\phi _{S3}
-\phi _{03}^S\left (\frac{2g_1}{3}\phi _{S2}+g_2'\xi '\right )+2g_3\xi _0\phi _{S1}\right |^2 \nonumber \\
&+2\left |-\frac{2g_1}{3}\phi _{01}^S\phi _{S3}+\phi _{02}^S\left (\frac{4g_1}{3}\phi _{S2}+g_2'\xi '\right )
-\phi _{03}^S\left (\frac{2g_1}{3}\phi _{S1}+g_2\xi \right )+2g_3\xi _0\phi _{S3}\right |^2 \nonumber \\
&+2\left |-\phi _{01}^S\left (\frac{2g_1}{3}\phi _{S2}-g_2'\xi '\right )-\phi _{02}^S\left (\frac{2g_1}{3}\phi _{S1}+g_2\xi \right )
+\frac{4g_1}{3}\phi _{03}^S\phi _{S3}+2g_3\xi _0\phi _{S2}\right |^2 \nonumber \\
&+2\left |-g_2\left (\phi _{01}^S\phi _{S1}+\phi _{02}^S\phi _{S3}+\phi _{03}^S\phi _{S2}\right )-2g_4\xi _0\xi \right |^2
+2\left |g_2'\left (\phi _{01}^S\phi _{S3}+\phi _{02}^S\phi _{S2}+\phi _{03}^S\phi _{S1}\right )\right |^2.
\end{align}
Taking VEVs and VEV alignments in Eqs.~(\ref{eq:alignment-vT}) and (\ref{eq:alignment-vS}), 
the scalar potential including driving fields of Eq.(\ref{eq:scalar-potential-driving}) are rewritten as
\begin{align}
\label{eq:scalar-potential-driving-2}
V_d^T&=2\left |M\phi _{01}^T\right |^2+8\left |M\phi _{03}^T\right |^2+8\left |M\phi _{02}^T\right |^2, \nonumber \\
V_d^S&=2\left |\left [\phi _{01}^S\left (\frac{4g_1}{3}c_S-g_2\right )-\frac{2g_1}{3}c_S\phi _{02}^S
-\phi _{03}^S\left (\frac{2g_1}{3}c_S+g_2\right )+2g_3c_S\xi _0\right ]u\right |^2 \nonumber \\
&+2\left |\left [-\frac{2g_1}{3}c_S\phi _{01}^S+\phi _{02}^S\left (\frac{4g_1}{3}c_S+g_2\right )
-\phi _{03}^S\left (\frac{2g_1}{3}c_S+g_2\right )+2g_3c_S\xi _0\right ]u\right |^2 \nonumber \\
&+2\left |\left [-\phi _{01}^S\left (\frac{2g_1}{3}c_S-g_2\right )-\phi _{02}^S\left (\frac{2g_1}{3}c_S+g_2\right )
+\frac{4g_1}{3}c_S\phi _{03}^S+2g_3c_S\xi _0\right ]u\right |^2 \nonumber \\
&+2\left |\left [-g_2c_S\left (\phi _{01}^S+\phi _{02}^S+\phi _{03}^S\right )-2g_4\xi _0\right ]u \right |^2
+2\left |g_2'c_S\left (\phi _{01}^S+\phi _{02}^S+\phi _{03}^S\right )u\right |^2,
\end{align}
where we define $c_S^2=g_4/(3g_3)$ and we eliminate trilinear coupling $g$ and VEV of flavon $v_T$, $v_S$, and $u'$ in Eqs.~(\ref{eq:alignment-vT}) and (\ref{eq:alignment-vS}).
Then, the VEVs of driving fields $\phi _0^T$, $\phi _0^S$, and $\xi _0$ are zeros which are derived from 
the condition of the potential minimum ($V_d^T=0$ and $V_d^S=0$) in Eq.~(\ref{eq:scalar-potential-driving-2}) such as trivial solution;
\begin{equation}
\label{eq:driving-fields-VEV}
\langle \phi _{0i}^T\rangle =\langle \phi _{0i}^S\rangle =\langle \xi _0\rangle =0.
\end{equation}
Therefore flavons take VEVs in Eqs.~(\ref{eq:alignment-vT}) and (\ref{eq:alignment-vS}), which break $A_4$ and driving fields take zero VEVs in Eq.(\ref{eq:driving-fields-VEV}).
Thus in $A_4$ breaking scale, Higgs $\mu $ term does not have mass scale such as $\mu =\lambda \langle \xi _0\rangle =0$. 
In the next subsection, we will discuss how to get non-zero $\mu $ term from soft SUSY breaking.

\subsection{Higgs $\mu $ term from soft SUSY breaking}
Let us discuss the non-zero Higgs $\mu $ term in this subsection.
After $A_4$ breaking, $\mu =\lambda \langle \xi _0\rangle $ is zero because VEV of driving field $\langle \xi _0\rangle =0$, which we discussed previous subsection.
Then, we consider soft SUSY breaking term in order to obtain Higgs $\mu $ term. 
The Lagrangian including soft SUSY breaking terms are written as
\begin{equation}
\mathcal{L}_\text{soft}\supset g_3A_{\phi _S}\xi _0\phi _S\phi _S -g_4A_{\xi }\xi _0\xi \xi +h.c.,
\end{equation}
where $A_{\phi _S}$ and $A_{\xi }$ are trilinear soft SUSY breaking A-terms and we assume that the A-terms are proportional to the trilinear couplings $g_3$ and $g_4$, respectively. 
Taking VEVs of flavons $\phi _S$ and $\xi $ as $\langle \phi _S\rangle =v_S(1,1,1)$ and $\langle \xi \rangle =u$, 
the Lagrangian including driving field $\xi _0$ is written as
\begin{align}
\mathcal{L}_{\xi _0}&\supset 3g_3A_{\phi _S}v_S^2\xi _0 -g_4A_{\xi }u^2\xi _0-12\left |g_3v_S\xi _0\right |^2-4\left |g_4u\xi _0\right |^2+h.c. \nonumber \\
&=\left (A_{\phi _S}-A_{\xi }\right )g_4u^2\xi _0-4|g_4|\left (|g_3|-|g_4|\right )|u\xi _0|^2+h.c..
\end{align}
Then after soft SUSY breaking, the driving field $\xi _0$ gets VEV as 
\begin{equation}
\langle \xi _0\rangle =\frac{\left (A_{\phi _S}-A_{\xi }\right )g_4u^2}{4|g_4|\left (|g_3|-|g_4|\right )|u|^2}\simeq \mathcal{O}(A_{\cancel{\text{SUSY}}})~\text{TeV},
\end{equation}
where $A_{\cancel{\text{SUSY}}}$ is soft SUSY breaking scale. Therefore we obtain Higgs $\mu$ term as
\begin{equation}
\mu =\lambda \langle \xi _0\rangle \simeq \mathcal{O}(A_{\cancel{\text{SUSY}}})~\text{TeV}.
\end{equation}

\section{Axion physics}
\label{sc:axion}

The $U(1)_{PQ}$ symmetry in our model gives rise to an axion field whose VEV cancels the QCD $\theta$ term and thereby solves the strong $CP$ problem~\cite{Peccei:1977hh}.
In the model, the two Higgs fields have $U(1)_{PQ}$ charges as in the DFSZ model \cite{dfsz} and the axion has properties similar to the DFSZ axion.
We in this section describe the formulas for the axion decay constant, axion mass and axion coupling with photons in the model and discuss their connection with the flavon sector.
Also, the current experimental bounds on axion properties are studied.

\subsection{Axion properties}

In our model, the $U(1)_{PQ}$ symmetry is spontaneously broken at the flavon VEV scale by the VEVs of fields $\phi_S, \, \xi, \, \xi^{\prime}$.
We hereafter assume that the flavon VEV scale is much higher than the electroweak scale.
The axion field $a$ is then given by a linear combination of the phase components of $\phi_S, \, \xi, \, \xi^{\prime}$,
\begin{align}
\phi_{Si} = (v_S + h_{Si}/\sqrt{2}) e^{i a_{Si}/(\sqrt{2}v_S)} \ (i=1,2,3), \nonumber \\
\xi = (u + h_{\xi}/\sqrt{2}) e^{i a_{\xi}/(\sqrt{2}u)}, \ \ \ \ \ \xi^{\prime} = (u^{\prime} + h_{\xi^{\prime}}/\sqrt{2}) e^{i a_{\xi^{\prime}}/(\sqrt{2}u^{\prime})},
\end{align}
in the following way:
\begin{align}
a &= \frac{ \sqrt{2}v_S (a_{S1} \, + \, a_{S2} \, + \, a_{S3}) \, + \, \sqrt{2}u a_{\xi} \, 
+ \, \sqrt{2}u^{\prime} a_{\xi^{\prime}} }{ \sqrt{6v_S^2 \, + \, 2u^2 \, + \, 2u^{\prime \, 2}} }.
\end{align}
The axion decay constant, $F_a$, is defined in the basis where the Dirac masses of all quarks and leptons are set to be real through axial phase transformations on fermions, 
$\psi \rightarrow e^{i \gamma_5 \alpha} \psi$. 
In this basis, the couplings of the axion to SM fields other than derivative interactions are given by
\begin{align}
{\cal L}_{{\rm axion \, couplings}} &= \frac{a}{F_a} \frac{g_C^2}{32\pi^2} G_{\mu \nu}^a \tilde{G}^{a \, \mu \nu} \ 
+ \ c_{a \gamma \gamma} \frac{a}{F_a} \frac{e^2}{32\pi^2} F_{\mu \nu} \tilde{F}^{\mu \nu},
\end{align}
where $g_C$ and $e$ are the QCD and electromagnetic coupling constants, respectively.
Here the axion decay constant $F_a$ is given by
\begin{align}
F_a &= \frac{ \sqrt{6v_S^2 \, + \, 2u^2 \, + \, 2u^{\prime \, 2}} }{ N_{DW} },
\end{align}
 where $N_{DW}$ is the domain wall number and $N_{DW}=2N_{qg}=6$ in the current model, with $N_{qg}$ denoting the number of quark generations.
It is remarkable that the axion decay constant is predicted to be of the same order as the flavon VEVs.
In the same basis, the ratio of the axion couplings to gluons and photons, $c_{a \gamma \gamma}$, is calculated by summing
 up-type quark, down-type quark and charged lepton contributions to the axion-photon coupling as
\begin{align}
c_{a \gamma \gamma } &= 2 \frac{ N_{qg} N_c q_u Q_u^2 + N_{qg}N_c(1-q_u)Q_d^2 + N_{lg}(1-q_u)Q_e^2 }{ N_{qg} } = \frac{8}{3},
\end{align}
where the number of colors is $N_c=3$, the number of lepton generations is $N_{lg}=3$, and the electric charges of up-type quarks, 
down type-quarks and charged leptons are $Q_u=2/3$, $Q_d=-1/3$ and $Q_e=-1$, respectively.

We next estimate the axion mass and the effective axion-photon coupling below the QCD confinement scale.
This is easily done by imposing the following axial phase transformations on the three light quarks $u, d, s$ to set the axion-gluon coupling to be zero:
\begin{align}
u &\rightarrow e^{i \gamma_5 \alpha_u} u, \ \ \ d \rightarrow e^{i \gamma_5 \alpha_d} d, \ \ \ s \rightarrow e^{i \gamma_5 \alpha_s} s \nonumber \\
&{\rm with \ \ \ } \frac{a}{F_a} - 2\alpha_u - 2\alpha_d - 2\alpha_s=0.
\label{axialtrans}
\end{align}
Note that $\alpha_u, \, \alpha_d, \, \alpha_s$ are not constants, but are dynamical fields that share the physical degree of freedom of the axion.
A convenient choice for $\alpha_u, \, \alpha_d, \, \alpha_s$ is 
\begin{align}
\alpha_u &= \frac{a}{2F_a}\frac{1}{1+z+w}, \ \ \ \alpha_d = \frac{a}{2F_a}\frac{z}{1+z+w}, \ \ \ \alpha_s = \frac{a}{2F_a}\frac{w}{1+z+w} \nonumber \\
&{\rm where \ \ \ } z \equiv \frac{m_u}{m_d}, \ \ \ w \equiv \frac{m_u}{m_s}.
\label{alphas}
\end{align}
Eq.~(\ref{alphas}) leads to $\alpha_u m_u = \alpha_d m_d = \alpha_s m_s$, with which the axion does not mix with $\pi^0$ and $\eta$ mesons in the chiral Lagrangian. 
With Eq.~(\ref{alphas}), the mass term for the axion, $\pi^0$ meson and $\eta$ meson in the chiral Lagrangian is given by 
(the mixing between $\eta$ and $\eta'$ mesons is neglected)
\begin{equation}
\label{axionmesonmass}
{\cal L}_{{\rm chiral}} = \frac{1}{2} v_{\chi}^3 {\rm tr}\{ \Sigma M + M^{\dagger} \Sigma^{\dagger} \} ,
\end{equation}
where 
\begin{align}
\Sigma &= \exp\left[2i \, {\rm diag}\left( \frac{\pi^0}{2f_{\pi}}+\frac{\eta}{2\sqrt{3} f_{\pi}}, 
\ -\frac{\pi^0}{2f_{\pi}}+\frac{\eta}{2\sqrt{3} f_{\pi}}, 
\ -\frac{\eta}{\sqrt{3}f_{\pi}} \right) \right ], \nonumber \\
M &= {\rm diag}(m_u e^{i2\alpha_u}, \, m_d e^{i2\alpha_d}, \, m_s e^{i2\alpha_s}),
\end{align}
and $v_{\chi}$ corresponds to the scale of chiral symmetry breaking.
Note that $M$ contains the physical degree of freedom of the axion through $\alpha_u, \, \alpha_d, \, \alpha_s$.
After diagonalizing the mass term Eq.~(\ref{axionmesonmass}), the axion mass, the $\pi^0$ mass and the $\eta$ mass are given by
\begin{align}
m_a^2 &= \frac{v_{\chi}^3 m_u}{F_a^2} \frac{1}{1+z+w}, \nonumber \\
m_{\pi^0}^2 &= \frac{v_{\chi}^3 m_u}{24 f_{\pi}^2 zw} \, \left[ \, 3(1+z)w + (4z+w+wz) \right. \nonumber \\
&- \left. \sqrt{ -48 zw(1+z+w) + \{ 3(1+z)w + (4z+w+wz) \}^2 } \, \right], \nonumber \\
m_{\eta}^2 &= \frac{v_{\chi}^3 m_u}{24 f_{\pi}^2 zw} \, \left[ \, 3 (1+z)w + (4z+w+wz) \right. \nonumber \\
&+ \left. \sqrt{ -48 zw(1+z+w) + \{ 3(1+z)w + (4z+w+wz) \}^2 } \, \right].
\label{mass}
\end{align}
Using Eq.~(\ref{mass}), we relate the axion mass with the $\pi^0$ mass in terms of $z = m_u/m_d$, $w = m_u/m_s$, $F_a$ and $f_{\pi}$ by eliminating $v_{\chi}^3 m_u$.
Substituting the experimental values $m_{\pi^0}=135$~MeV, $f_{\pi}=93$~MeV, $z=0.56$, and $w=0.028$~\cite{Agashe:2014kda}, we obtain
\begin{align}
m_a &= 6.0 \times 10^{-6} \, {\rm eV} \left( \frac{10^{12} \, {\rm GeV}}{F_a} \right).
\label{axionmass}
\end{align}
In the same basis, the axion coupling to photons is given by
\begin{align}
{\cal L}_{{\rm axion \, couplings}} &= \left\{ c_{a \gamma \gamma} - 2 \frac{ 2\alpha_u N_c q_u Q_u^2 + (2\alpha_d+2\alpha_s) N_c(1-q_u)Q_d^2 }{ N_{qg} } \right\}
 \frac{a}{F_a} \frac{e^2}{32\pi^2} F_{\mu \nu} \tilde{F}^{\mu \nu} \nonumber \\
 &= \left( \frac{8}{3} - \frac{2}{3} \frac{4+z+w}{1+z+w} \right) \frac{a}{F_a} \frac{e^2}{32\pi^2} F_{\mu \nu} \tilde{F}^{\mu \nu}.
\label{axionphoton}
\end{align}

\subsection{Experimental constraints}

The observation of globular-cluster stars puts a bound on the axion-photon coupling.
Defining the axion-photon coupling $g_{a\gamma \gamma}$ by ${\cal L} =(1/4)g_{a\gamma \gamma} a F_{\mu \nu}\tilde{F}^{\mu \nu}$,
 we find that the bound is expressed as~\cite{Ayala:2014pea}
\begin{align}
\vert g_{a\gamma \gamma} \vert &< 6.6\times 10^{-11}~{\rm GeV}^{-1}, 
\label{eq:globular}
\end{align}
at $95\% \ {\rm C.L.}$. 
The cooling of supernova 1987A puts a bound on the axion decay constant $F_a$ as~\cite{1987a}
\begin{align}
F_a &\gtrsim 3.7 \times 10^8~{\rm GeV}.
\label{eq:1987a}
\end{align}
Eq.~(\ref{eq:1987a}) gives a more severe bound than Eq.~(\ref{eq:globular}) on the axion decay constant.
In the model, this also gives a lower bound of the flavon VEV scale and the right-handed Majorana neutrino mass scale. 
For example, if $v_S=u=u^{\prime}$ and the Majorana Yukawa coupling is $O(0.1)$, the right-handed Majorana neutrino mass $M_R$ is bounded from below as
\begin{align}
M_R &\gtrsim 10^8~{\rm GeV}.
\label{majorana}
\end{align}

If the axion is the dominant constituent of the dark matter, further experimental constraints apply.
The ADMX Collaboration reports a bound on the ratio of the axion mass and the axion-photon coupling in the axion dark matter scenario \cite{admx},
 but the DFSZ axion is not yet constrained by this experiment. 
Since the ratio of the axion mass and the axion-photon coupling is the same in our model and in the DFSZ model, our model also evades the ADMX bound.

\newpage 

\section{Summary}
\label{sc:summary}
We have constructed a model based on the $A_4 \times U(1)_{PQ}$ flavor symmetry.
The $A_4$ symmetry induces TBM in the lepton sector to explain large mixing angles, while it is consistent with the quark sector of the MSSM, since all the quark fields are assigned to $A_4$ trivial singlets.  
On the other hands, the $U(1)_{PQ}$ symmetry is introduced to forbid unwanted couplings in the lepton sector and to solve the strong $CP$ problem.
In the model, we have introduced the $A_4$ triplet and singlet flavons with $U(1)_{PQ}$ charges to break the symmetry and to obtain observed masses and mixing angles in the lepton sector. 
Furthermore, Majorana mass of right-handed neutrinos is generated after the symmetry breaking by flavon VEVs.

We have analyzed the scalar potential including flavons and driving fields in the framework with SUSY and $U(1)_R$ symmetry.
We have shown that the alignment of flavon VEVs is relevant to obtain the correct masses and mixing angles in the lepton sector.
In addition, an additional $A_4$ singlet-prime flavon breaks TBM to obtain non-zero $\theta_{13}$.
Moreover non-zero Higgs $\mu$ term can be obtained from soft SUSY breaking term which induces non-zero VEV of a driving field $\xi_0$.
As a result, $\mu$ term is given by the VEV of the driving field which is at the soft SUSY breaking scale.

We have studied the axion in the model, which has properties similar to the DFSZ axion, since our Higgs fields have $U(1)_{PQ}$ charges.
The axion field $a$ is given by a linear combination of the phase components of flavons which have non-zero VEVs, and the axion decay constant $F_a$ is obtained in terms of the VEVs of flavons.
Remarkably, the axion decay constant is of the same order as the flavon VEVs in the model.
We have also derived the axion mass and axion-photon coupling as functions of the axion decay constant. 
As to experimental constraints on the axion decay constant, it should be larger than $O(10^8)$ GeV due to the constraint from cooling of supernova 1987A,
 and this can be interpreted as a lower bound on the flavon VEV scale in the model.
We note that the Majorana mass of right-handed neutrino $\nu_R$ is related to $F_a$ in our model since both of them are determined by the flavon VEVs.
The Majorana mass should satisfy $M_R \gtrsim 10^8$ GeV when we assume that the Yukawa coupling for $\nu_R$ is $O(0.1)$ and that all flavon VEVs are of the same order.
The axion in our model can be the dominant constituent of dark matter. In such a case, still, the axion is not constrained by the ADMX experiment, just like the DFSZ axion.

\vspace{0.5cm}
\noindent
{\bf Acknowledgement}

YS is supported by Grant-in-Aid for JSPS Fellows No.16J05332.
This work is also supported in part by the National Research Foundation of Korea (NRF) Research Grant NRF-2015R1A2A1A05001869 (YS and TY). 

\newpage

\appendix 

\section{Multiplication rule of $A_4$ group}
\label{sec:multiplication-rule}
We show the multiplication of $A_4$ group.
The multiplication rule of the triplet is 
written as follows,
\begin{eqnarray}
&&
\left (
\begin{array}{c}
a_1\\
a_2\\
a_3
\end{array}
\right )_{\bf 3}
\otimes 
\left (
\begin{array}{c}
b_1\\
b_2\\
b_3
\end{array}
\right )_{\bf 3}
=\left (a_1b_1+a_2b_3+a_3b_2\right )_{\bf 1} \nonumber \\
&\oplus &\left (a_3b_3+a_1b_2+a_2b_1\right )_{{\bf 1}'}
\oplus \left (a_2b_2+a_1b_3+a_3b_1\right )_{{\bf 1}''} \nonumber \\
&\oplus &\frac{1}{3}
\left (
\begin{array}{c}
2a_1b_1-a_2b_3-a_3b_2 \\
2a_3b_3-a_1b_2-a_2b_1 \\
2a_2b_2-a_1b_3-a_3b_1
\end{array}
\right )_{\bf 3} \nonumber \\
&\oplus &\frac{1}{2}
\left (
\begin{array}{c}
a_2b_3-a_3b_2 \\
a_1b_2-a_2b_1 \\
a_3b_1-a_1b_3
\end{array}
\right )_{\bf 3}\ .
\end{eqnarray}
More details are shown in the review~\cite{Ishimori:2010au}-\cite{King:2014nza}.


\end{document}